\documentclass[preprint]{aastex}
\def\simgr{\,\hbox{\hbox{$ > $}\kern -0.8em \lower 1.0ex\hbox{$\sim$}}\,}
\def\simle{\,\hbox{\hbox{$ < $}\kern -0.8em \lower 1.0ex\hbox{$\sim$}}\,}

\shortauthors{THORSTENSEN \& ARMSTRONG}
\shorttitle{FIRST J102347.6+003841}
\begin{document}
\title{Is  FIRST J102347.6+003841 Really a Cataclysmic Binary?
\footnote{Based on observations obtained at the MDM Observatory, operated by
Dartmouth College, Columbia University, Ohio State University, and
the University of Michigan.}
}

\author{John R. Thorstensen}
\affil{Department of Physics and Astronomy\\
6127 Wilder Laboratory, Dartmouth College\\
Hanover, NH 03755-3528;\\
john.thorstensen@dartmouth.edu}

\author{Eve Armstrong}
\affil{Department of Astronomy\\ 
Columbia University\\ 
550 West 120th Street\\ 
New York, NY 10027}

\begin{abstract}
The radio source FIRST J102347.6+003841 was presented as
the first radio-selected
cataclysmic variable star.  In the discovery paper, \citet{bond}
show a spectrum consistent with a magnetic AM Her-type system,
or polar, featuring strong Balmer lines, 
He I, and He II emission lines, and a light curve
with rapid, irregular flickering. In contrast, \citet{woudtwarner}
found a smoothly-varying light curve with a period near
4.75 h and one minimum per orbit, indicating that the
state of the system had changed dramatically.  
We present time-resolved spectra showing a 
superficially normal, mid-G type photosphere, with no detectable 
emission lines.  The absorption-line radial velocity 
varies sinusoidally, with semi-amplitude $268 \pm 4$ km s$^{-1}$,
on the orbital period, which
is refined to 0.198094(2) d.  At this orbital period the
secondary's spectral type is atypically early, suggesting 
an unusual evolutionary history.
We also obtained photometry around the orbit in 
$B$, $V$, and $I$.  The light curve resembles 
that observed by \citet{woudtwarner}, and the colors
are modulated in a manner consistent with a heating effect.
A simple illumination model matches the observations strikingly 
well with a Roche-lobe filling 
secondary near $T_{\rm eff} = 5650$ kelvin being
illuminated by a primary with an isotropic luminosity
of $\sim 2$ L$_{\odot}$.  The modest amplitude of the observed
modulation constrains the orbital inclination $i$
$\sim 55$ degrees or less, unless the gravity 
darkening is artificially reduced.  Combining the low $i$ with the
secondary's velocity amplitude gives a primary star
mass above the Chandrasekhar limit when conventional
gravity darkening is assumed.  We consider
the robustness of this conclusion, and examine the possibility
that the compact object in this system is not a white dwarf, 
in which case this is not actually a cataclysmic variable.
On close examination, FIRST J102347.6+003841 defies easy classification.
%
\end{abstract}

\section{Introduction}

Cataclysmic variable stars (CVs) are close binaries in which a white
dwarf accretes matter from a less-evolved companion (the secondary),
which usually resembles a lower main sequence star.  CVs have a rich
phenomenology, and the theory of CVs ranges across wide areas of
astrophysics; \citet{warn} presents a comprehensive review. 

CVs draw attention to themselves through a variety of channels,
most importantly through variability (especially for novae
and dwarf novae), X-ray flux, unusual optical colors, or emission-line
spectra.  Magnetic CVs are sometimes weak radio sources, and 
the subject of this paper, FIRST J102347.6+003841 (hereafter 
FIRST 1023+00), is thought to be the only CV that was
initially found from its radio flux.  \citet{bond} 
identified the radio source with a 17-th magnitude, flickering
star, the spectrum of which showed emission lines
typical of an AM-Her class CV, or `polar'.  The object was
independently discovered in the 
Sloan Digital Sky Survey; \citet{szkodysdss2} show
a spectrum with a strong blue continuum and high-excitation
emission lines.   They remark that the lines are not as strong as in 
typical AM Her stars (`polars'), but suggested that the
system might be a DQ Her star (or `intermediate polar').  
\citet{woudtwarner} obtained photometry showing a smooth
light curve with a period of 4.75 h, in sharp contrast to the
flickering seen by \citet{bond}.  They attribute the
modulation to a heating (sometimes called `reflection') effect.
The very different photometric behavior seen by 
\citet{woudtwarner} suggests that FIRST 1023+00 undergoes
dramatic changes in state.  The SDSS data already suggest
such a conclusion, as the broadband colors are rather red
(Table 1 of \citealt{szkodysdss2}), but the spectrum (which
was not simultaneous with the photometry) appears quite blue
blue (Fig.~2 of their paper).

Here we present spectra which show only late-type absorption
features, dramatically different from the spectra published
by \citet{bond} and \citet{szkodysdss2}.  The radial
velocity of the late-type spectrum is modulated on the the 4.75-h
photometric period, which allows us to refine the ephemeris and
constrain the dynamics.  The spectral type and color of the
late-type star is atypically early for the orbital period, 
suggesting that it began mass
transfer after significant nuclear evolution had taken place.  We
also present time-resolved $BVI$ photometry, which shows an orbital
color modulation consistent with a heating effect.  We model this
effect quantitatively in the Roche geometry.  These models,
together with the completely atypical quiescent behavior of this
system, suggest a radically different interpretation of the 
object, in which the compact object is a neutron star or a black
hole.

\section{Observations}

Table 1 contains a journal of the observations.  A single survey
spectrum was taken 2003 January, which showed only absorption
lines.  We initially thought we had observed the wrong star.  Further
observations in 2004 January showed that there was no error in
the identification, and revealed the periodicity in the
absorption velocity.  We obtained more spectroscopy in 2004
March, but poor weather limited the coverage.  Also in 2004 March
we obtained photometry and some differential time-series
photometry.  In 2004 May, near the end of the observing season,
we obtained differential time-series $BVI$ filter photometry
covering the entire orbit.

\subsection{Spectroscopy}

The spectroscopic protocols were for the most part identical to those
described in \citet{longp03}. 
Briefly, we used the `modular'
spectrograph, a 600 line mm$^{-1}$ grating, and a SITe 2048$^2$ CCD.
This yielded 2 \AA\ pixel$^{-1}$ from 4210 to 7560 \AA\ (vignetting
severely toward the ends), and typical resolution of 3.5 \AA\ FWHM.  The
spectra were flux-calibrated using observations of standard stars, but
spectrophotometry was approximate because of uncalibratable losses at
the the 1-arcsecond slit, occasional cloud, and an instrumental effect
of uncertain origin, which sometimes introduces wavelike distortions in
stellar continua.  

For absorption-line radial velocity measurements we used the
cross-correlation method of \citet{tonrydavis}, as implemented by
\citet{kurtzmink} in the {\it rvsao} package.  The
cross-correlation template was a composite of many IAU velocity
standard spectra, as described in \citet{longp03}.  



\subsection{Photometry}

In 2004 March we obtained direct images, including three contiguous
sets of $UBVI$ images, with the Hiltner 2.4m telescope and a SITe
$2048^2$ CCD detector which gave 0.275 arcsec pixel$^{-1}$.  We
derived standard magnitudes from observations of \citet{landolt92}
standard-star fields; Table 2 gives the results.  The magnitude
transformations had standard deviations $< 0.02$ mag for $V$, $B-V$,
and $V-I$, but $0.08 $ mag for $U-B$ (which often gives large
scatter).  For the brightest stars the magnitudes and colors from the
three image sets agreed to $\sim 0.01$ mag, indicating that conditions
were clear.  Fig.~1 shows the stars measured and their magnitudes.  In
addition to the standardized photometry, we took 28 $I$-band
`snapshots' over two nights in order sample the light curve and
establish the relationship between photometric and spectroscopic
phase. These were reduced using the IRAF implementation of DAOPHOT in
a semi-automated mode.  The differential magnitudes  were converted
to approximate standard $I$ magnitudes using the photometric sequence.  

We obtained more extensive time-series photometry of the object on three
nights in 2004 May using the 1.3 m McGraw-Hill telescope and a
thinned SITe $1024^2$ CCD.  To minimize dead time we binned the
readout $2\times2$ and cropped the area to yield a $256^2$ image
with a scale of 1.018 arcsec per (binned) pixel.  We cycled
through 120 s exposures in the $B$ filter, and 60 s exposures in
$V$ and $I$ filters.  After bias subtraction and flat fielding
using twilight sky exposures, we measured magnitudes in 4-arcsec
diameter apertures using IRAF.  Differential magnitudes were
computed relative to the $V = 14.86$ comparison star west of the
variable (see Fig.~1).  The $V = 17.40$ mag star to the southeast
served as a check; typical precisions were $< 0.02$ mag (after we
discarded a few images taken through intermittent heavy cloud).
Hour angle constraints prevented us from covering the entire
orbit in one night, but we were able to piece together coverage
of the whole orbit from the three nights' data.   When phase
redundancy was obtained the different nights agreed closely --
there was no trace of secular variability.  The differential
magnitudes and colors were standardized using the sequence in
Table 2.

\section{Results and Analysis}

Fig.~2 shows the mean spectrum, shifted to the rest frame of the
secondary using the ephemeris described later.  It is essentially
that of a late-type star, with no emission lines.  The flux level
implies $V = 17.5$, in good agreement with the filter photometry.

Table 3 lists the cross-correlation radial velocities.  A
search of these yields a strong periodicity near 4.75 hr, but
with several possible precise periods corresponding to different
choices of cycle count over the 50-day interval between 2004
January and March.  Fortunately, the photometric period published
by \citet{woudtwarner} is consistent with only one of the precise
periods.  Once the 50-day cycle count is decided,  the ephemeris
extends unambiguously back to the single 2003 January point,
further tightening the period.  To help us verify the ephemeris,
P. Woudt kindly sent the time-series photometry data used by
\citet{woudtwarner}; combining this with our own time-series
photometry again yielded the same period, confirming the
ephemeris.  The period found here differs slightly from that
found by \citet{woudtwarner} from their photometric data alone;
our period supersedes theirs, since it is based on a longer time
span and has greater redundancy.  Table 4 gives the parameters
of the best-fitting sinusoid, and Fig.~3 shows the folded
velocities with the sinusoid superposed.

Fig.~4 shows the 1.3m time series photometry folded on the
spectroscopic ephemeris, in which phase zero corresponds to the
inferior conjunction of the secondary star.  The light curve is
similar to that found by \citet{woudtwarner}.  Minimum light
corresponds closely with inferior conjunction, as expected for 
a heating effect.
The $B-V$ and $V-I$ colors are a also modulated, with the star
bluest at maximum light.  The curves in Fig.~4 are from a
model described later. 

We rectified all the spectra and created the phase-averaged
greyscale representation shown in Fig.~5.  The most
remarkable feature is the complete absence of {\it any} emission
lines.  Greyscale representations of this kind can sometimes bring
out subtle spectral features, but this looks remarkably ordinary.
There is also no strong modulation of the spectrum around
the orbit.  The coverage is unfortunately nonuniform (refer
to Fig.~3); for a portion of the figure around phase 1,
only one observation contributes, so the data remain
unchanged with phase.  However, coverage is densest
on the descending side of the velocity curve, which 
corresponds to superior conjunction of the secondary.
This is the portion of the orbit where the heated face of
the secondary is turned toward us, and we still do not 
see any emission lines.

To estimate the mean spectral type of the secondary we used the
spectral decomposition procedure outlined in \citet{longp03},
basically subtracting scaled versions of spectral-type standard
stars away from the mean spectrum, and looking for good
cancellation of the late-type features.  The K- and M-dwarf
library spectra taken with the same instrument proved to be too
cool to match, so we used G-type library spectra from
\citet{jacoby}.  These matched reasonably well around mid-G,
but with a significant mismatch at the sodium D lines, which are
stronger than they should be for the spectral type.  The
decomposition was consistent with {\it all} the light coming from
the secondary star -- there was no need to include any
contribution from a hot component.  This conclusion is
corroborated by the colors in Table 2, and the $V-H$ and $V-J$
colors found by combining the present data with the $J$ and $H$
magnitudes from the 2MASS All Sky Data Release.  \citet{pickles}
tabulates normal G-dwarf colors in many passbands, and the
observed colors are all consistent with a type between G5 and G7.
Remarkably, this is even true for the $U$-band, in which one
might expect to start seeing a hot component if one were present.
Although the phase-averaged spectrum is not sensitive enough to
show spectral-type variation, the color modulation corresponds to
a half-amplitude of around 3 subtypes, corresponding to $\sim
180$ kelvin around mid-G.

\section{A Light Curve Modeling Program}

The remarkable light and color curve of this system motivated us to
construct a light-curve modeling program.  In our model, the
secondary fills its Roche critical lobe, as CV secondaries generally
do; although mass transfer was not evident during our observations,
the secondary should still nearly fill its Roche lobe, since the
system was actively transferring mass within the last few years and
the Kelvin-Helmholtz relaxation time of the secondary's envelope
should be much longer than that.  

To tessellate the model secondary, we used an icosahedron-based
geodesic dome construction similar to (but less elaborate than)
that advocated by \citet{hendrygeod}.  Each face of the
icosahedron was divided into 25 sub-triangles, and the vertices
were then `pushed out' to the Roche surface, to form a 500-sided
polyhedron.   A vertex of the icosahedron was oriented toward the
primary, so that the `pointy' part of the Roche lobe near the
inner Lagrangian point was modeled adequately.  Fig.~6 shows the
geometry used for the model secondary.  

We define the `base' effective temperature of the secondary, $T_{\rm
eff,0}$, as the effective temperature in the absence of gravity
darkening or heating effects.  The local gravity $g$ at each segment
was computed (at the segment's center) and used to determine the
gravity darkening through $(T/T_{\rm eff,0}) = (g/\bar g)^\alpha$.
For our standard models we used $\alpha = 0.08$, as recommended 
by \citet{lucy} for stars with convective envelopes.  For limb-darkening we 
used a simple linear law, $I(\mu)/I_0 = 1 - x (1  -\mu)$, where 
$\mu$ is the cosine
of the angle away from the local vertical (as is standard), 
and $x$ is the linear limb-darkening
coefficient, which we took to be 0.5 for our standard models.  More
elaborate limb-darkening laws can be justified \citep{vanhamme}
but we chose this for simplicity. 
To explore the effects of limb- and gravity-darkening, we also
computed models with $\alpha$ = 0 and $\alpha = 0.16$, and 
models with $x = 0$ and $x = 0.75$.  

The radiation illuminating the model secondary is assumed to come
from the primary star's location, and the primary is assumed to act
as a point source.  The incident radiation was assumed to thermalize
completely and be re-radiated in a manner similar to a normal
stellar atmosphere, so that the effect of the irradiation could be
adequately treated as an increase in the local effective temperature
sufficient to carry the added luminosity.
This assumption may be questionable, but in the absence of any
information about the spectrum of the incident radiation it seemed
to be a good starting point.

Once the effective temperature of each segment was determined,
the surface brightness in $V$ radiated perpendicular to each
segment was estimated based on stellar atmosphere models.
More specifically, we computed the quantity
$S_V$, which we define as the absolute $V$ magnitude of a 
1 R$_{\odot}$ star of the same surface brightness; this is
$$S_V = 42.367 - {\rm BC}(V) - 10 \log T_{\rm eff},$$
where BC$(V)$ is the bolometric correction in the $V$ band.
\citet{bessellbc} give a convenient tabulation of BC$(V)$ 
versus $T_{\rm eff}$ for ATLAS9 models computed using 
Kurucz' codes; we created polynomial approximations to the
$S_V$, $\log T_{\rm eff}$ relation, and similarly for
$B-V$ and $V-I$ (which are also tabulated).  
Once $S_V$ was determined, the specific intensity 
normal to the surface was determined using the 
conversion factor 
$$I_{0V} = {2.346 \times 10^{8} \hbox{ erg s}^{-1} 
\hbox{ cm}^{-2} \hbox{ \AA}^{-1} \hbox{ sterad}^{-1} \over 1 - x/3},$$
for a 1 R$_{\odot}$ star with $V = 0$.  
This factor is based on 
(a) a flux $3.75 \times 10^{-9}$ erg cm$^{-2}$ s$^{-1}$
\AA $^{-1}$ in the $V$ band for a $V = 0$ star, (b) the solar radius
($6.96 \times 10^{10}$ cm), and (c) the conversion from
total monochromatic energy in the outgoing direction to 
specific intensity, which introduces factors of
$\pi$ and $1 - x/3$ from integration over solid angle.

After the perpendicular surface brightness of each surface element
had been determined, an orbital inclination $i$ was chosen, 
and the light and color curves were
computed by integrating over all elements visible to the observer
at each observed phase.  The result was the specific intensity times the
projected area, which was converted to a physical flux at a
fiducial distance by dividing by the square of that distance.
The assumed distance to the system was then adjusted to give
the best match between the theoretical and observed light curves.  
After some experimentation with interactive fitting, we constructed 
a program which automatically converges the light-curve fits by
adjusting three parameters, namely $i$, $T_{\rm eff,0}$,
and the luminosity of the primary $L_1$.  The
convergence was controlled by an inefficient but robust 
`brute-force' technique which thoroughly surveyed the surrounding
parameter space at each iteration.  The masses were chosen at the
start of each calculation and left
fixed during the optimization.

Before fitting the data, we corrected them for a reddening
$E(B-V) = 0.045$ mag, estimated by \citet{schlegel98}.
The $V-I$ colors were corrected using $E(V-I)/E(B-V) = 1.35$,
a value taken from \citet{hereddening}.  Including these
corrections raised $T_{\rm eff,0}$ by $\sim 150$ kelvin.

To check the program, we used results from \citet{bochkarev}, who
computed light-curve amplitudes  for Roche-lobe filling
secondaries.  We adjusted the limb- and gravity-darkening
parameters in our program to match their choices, turned off the
external illumination, and used a simplified version of the
surface-brightness procedures to match their parameterization.
With these modifications, our code reproduced their tabulated
light-curve amplitudes to within a few thousandths of a
magnitude.  

For each model, the secondary's velocity amplitude $K_2$ 
followed from the fixed masses, the optimized inclination, and 
the known orbital period.  However, the amplitude could
be affected by non-uniform contributions to the line spectrum
across the secondary.  To establish a crude upper limit
to this effect, we also computed a quantity $K_2({\rm back})$, 
a radial velocity amplitude which included only contributions
from surface elements that were not externally illuminated.
The weight of each element in the velocity average was proportional
to its contribution to the light.

\section{Results of the Modeling}

We computed optimized light curves for primary masses $M_1$ of 
1., 1.4, 2., and 2.8 M$_{\odot}$, and secondary masses $M_1$ of 0.1, 0.2, 0.3,
0.5, 0.7, 0.9, and 1.2 M$_{\odot}$, constraining  
$M_2$ to be less than $M_1$.  (Exploratory models
with smaller $M_1$ did not come close to matching the observed
$K_2$, so we did not explore smaller values of $M_1$ systematically.)   
All these combinations gave similarly good fits to the light and color curves, 
the curves in Fig.~4 being typical.  From the good fits 
we conclude (a) that the model appears basically valid, and
(b) that the photometry alone carries essentially no information on the 
masses.  Table 5 gives a sampling of model fits; they are 
chosen to give either a predicted $K_2$ or $K_2$(back) within three
standard deviations of the observed value.

Other features of the results are:
\newcounter{Lcount}
\begin{list}{\arabic{Lcount})}{\usecounter{Lcount}}
\item The secondary's $T_{\rm eff,0}$ is constrained to a very narrow
range, the extremes in all the models being 5570 and 5740 kelvin.
This is consistent with the spectral type.

\item The inferred distance is almost entirely determined
by the secondary's assumed mass, because for a fixed $P_{\rm orb}$ in the
Roche geometry, the secondary star's radius 
$R_2 \sim M_2^{1/3}$, and the
surface brightness is fixed by the color.  The distance can be expressed as 
2.2 kpc $\times (M_2 / {\rm M_{\odot}}) ^ {1/3}$.

\item $L_1$ varies over a rather limited range from 1.5 to 4 L$_{\odot}$.
It is affected most strongly by $M_1$, because the binary separation 
increases with $M_1$ at a constant $P_{\rm orb}$, forcing the 
primary to shine more brightly to maintain a given heating effect at the
secondary's face.

\item For a fixed $M_2$, $K_2({\rm back})$ exceeds $K_2$ by a nearly 
constant amount.  At $M_2 = 0.7\ {\rm M}_{\odot}$, the difference
is $\sim 25$ km s$^{-1}$, while for  $M_2 = 0.1\ {\rm M}_{\odot}$
the difference drops to $\sim 15$ km s$^{-1}$.  

\item $K_2$ is almost independent of $M_2$.

\item $K_2$ is nearly independent of the choice of limb-darkening
coefficient, but is strongly affected by gravity darkening, in the
sense that increased gravity darkening decreases $K_2$.
The stronger the gravity darkening, 
the darker the back side of the secondary becomes, which results in 
a more face-on inclination, and hence a lower $K_2$.

\item The only models that come close to matching the 
observed $K_2 = 268 \pm 4$ km s$^{-1}$ have large values of $M_1$.  
The models with $M_1 = 1\ {\rm M}_{\odot}$ can only be made to fit if 
the gravity-darkening is reduced to zero {\it and} the observed $K_2$ is 
identified with $K_2({\rm back})$.  With the
standard assumptions, models with $M_1$ at or below the Chandrasekhar
limit fit the observed light curves poorly.  Fig.~7 illustrates this;
it shows $K_2$(back) as a function of $M_1$ for a run of models with 
$M_2 = 0.7\ {\rm M}_{\odot}$.

\end{list}

The last item is obviously provocative, but
the large value of $L_1$ is also unexpected and suggestive.  A luminosity
of $\sim 2$ L$_{\odot}$ is very high for a white dwarf and implies an 
effective temperature $\sim 50000$ kelvin \citep{bergeron}.  At this 
temperature there should be a strong ionizing flux, and in other close binaries 
harboring such hot white dwarfs this creates chromospheric
emission from the illuminated face of the secondary.  To estimate the magnitude
of the effect in this system, we scaled the results from the 
0.7-day WD+M3V binary EUVE 2013+400, in which the 
white dwarf is $\sim 50000$ kelvin \citep{thorstensen94}, and
found that a similar white dwarf in this system would 
create emission at H$\alpha$ with $\sim 3$\ \AA\ equivalent width (EW).  
Our observations show no hint of emission; the easily
detected H$\alpha$ absorption line has an EW of $\sim 2$ \AA. 
In any case, the white dwarfs observed in magnetic
cataclysmics are typically much cooler than this, averaging
only $\sim 16000$ kelvin \citep{sion99}.  On both these
grounds it seems unlikely that a normal white dwarf is 
powering the heating effect.

The implied mass is large because the heating model requires
a low inclination.
One natural suggestion for increasing $i$, hence lowering $M_1$,
is to add a source of steady light to the system, as would be
expected from the primary.  However, as noted earlier, the observed
colors from the near ultraviolet to the mid-infrared are all nicely
matched by pure G-star atmospheres.  We tried adding light
artificially, and of course this increased the inclination, but
unless the added light was also tuned to be almost exactly the color
of the secondary, the added light destroyed the color match.  In
particular, adding any significant amount of light similar in color
to a hot white dwarf completely destroyed the agreement between the
observed and modeled colors.

\section{Discussion}

What are we to make of this star?  Except for the unusual
discovery channel, the evidence available to \citet{bond} showed
a typical AM Her-type cataclysmic binary, yet since then it has
shown a very different face, most likely because mass transfer
has ceased, at least for the time being.
There is nothing in the minimum-light spectrum to suggest it is a 
cataclysmic binary; in particular there are no
emission lines.  Even at deep minimum, AM Her stars
such as EF Eri still show emission (see, e.g., \citealt{wheatley98}).

The early spectral type of the secondary is also unusual.
\citet{beuermann2ndry} compiled the spectral types of CV
secondaries as a function of P$_{\rm orb}$, and found that for
orbital periods $P_{\rm orb} < 3$ hr, nearly all were consistent
with main-sequence expectations, while for $P_{\rm orb} > 3$ hr,
secondaries were either near the main sequence or {\it cooler}
than expected.  At the 4.75 hr period of FIRST 1023+0038, the
spectral type of a secondary obeying the main-sequence
mass/radius/spectral-type relation would be near M2, {\it much}
cooler than observed, so FIRST 1023+0038 lies on the `wrong' side
of the trend line.  Recently \citet{thorstensenei} and
\citet{thorstensenqz} drew attention to two shorter-period
systems that also show anomalously warm secondaries, namely EI
UMa ($P = 64$ min) and QZ Ser ($P = 2.0$ hr).  The explanation
advanced in those papers was that these objects arose from
systems in which mass transfer began after the secondary had
undergone some nuclear evolution.  In QZ Ser, this is
corroborated by an apparent enhancement of the Na abundance,
since Na is bred out of Ne at temperatures consistent with H
burning in the CNO cycle.  The early spectral type and strong NaD
line in FIRST 1023+0038 suggest that the secondary has undergone
some nuclear evolution in this system as well.  The \citet{bond}
spectrum does show H emission, so H-burning evidently did not
complete in the material that is presently exposed on the
secondary's surface.

The lack of emission lines and the excellent fit to a pure
heating-effect light curve at low inclination presents the most
severe challenges to understanding.  Here are four scenarios which
might explain the observations.

{\it Malevolent Starspots.}  The constraints from the 
light-curve model might be relaxed or removed altogether if
there were extensive starspots with just the right 
distribution to mimic the observations.  
Starspots can cause strong secular variations in the 
light curves of rapidly-rotating, late-type stars; the 
secondary in V471 Tau \citep{ibanoglu} is a particularly striking
example.  We have not attempted to model this effect here, 
as the parameter space becomes
dauntingly large.  Starspots may play some role; the 
light curves shown by \citet{woudtwarner}, which were
taken about a year before ours, show an asymmetry which
we do not observe.  As a main driver for the light curve,
starspots are unattractive because they would have to
be fortuitously arranged to create a false heating effect
in both the light and color curves.  

{\it Residual heat.} If the heat absorbed by the secondary
were retained and re-emitted over a long time scale, one could
get by with a much less luminous primary.  We think this is
unlikely, because externally-irradiated stellar atmoshperes
are observed to respond rapidly to changes in the irradiating
flux.  In HZ Herculis, for example, the light curve shows
complex modulations in synchrony with the 35-day disk
precession period \citep{gerend}.

{\it A Triple System.}  If there were another late-type star
in the system, with nearly the same color as the observed
secondary, it would allow the inclination to be higher
and the inferred secondary mass to be lower.  However, there
is no evidence for a stationary line system, so we 
regarded this as unlikely.  Such a contribution might lurk
below our spectral resolution, but mixing in 
G-star light at the systemic velocity would be increase
the true $K_2$ still further.

{\it A massive, invisible primary.}  The data are naturally fit
by an invisible massive primary which emits about 2 L$_{\odot}$,
and the absence of emission lines from the secondary's irradiated
atmosphere suggests that the intercepted radiation is thermalized
deep within the secondary's atmosphere.  The system does not
resemble any known cataclysmic at minimum light.  Thus the {\it prima
facie} evidence suggests that the primary is not a white dwarf, and hence 
FIRST 1023+00 is not a cataclysmic binary.
If the primary isn't a white dwarf, what is it?

One attractive explanation is that the primary is a neutron star, so
the system is similar to a low-mass X-ray binary.  The strong
$\lambda 4686$ line in the \cite{bond} high-state spectrum is
consistent with an X-ray binary.  However, the lack of any known
X-ray source in the vicinity presents a vexing problem.  The duty
cycle for active mass transfer could be short, but at the
inferred low inclination and modest distance (for an X-ray binary)
it is difficult to see how any substantial
X-ray source could have remained unnoticed.  
On the positive side, transient X-ray sources are often
observed to have quiescent X-ray luminosities comparable
to that inferred for the primary star 
\citep{menou}.

A black-hole primary would be even more exotic, and in this
case the large velocity amplitude and low inferred inclination
would be explained very naturally.  Getting $\sim 2$ L$_{\odot}$
out of the compact object without creating any emission lines
could be problematic in that case.

The system does vaguely resemble V616 Mon \citep{gelino}, but in 
that case the secondary does not show a clear heating effect.
It is interesting that the primary in V616 Mon is evidently
a black hole, which `turns off' almost entirely in quiescence,
while neutron-star transients generally continue to emit at a
low level after they fade \citep{menou}.

To summarize, there does not seem to be any simple, plausible model
for this object that does not present further difficulties.  
Continued optical monitoring and a sensitive search for X-ray emission
might help resolve these issues.

{\it Acknowledgments.} The NSF supported this work through grants AST
9987334 and AST 0307413.  We thank the MDM staff for their support.
Our special thanks go to Patrick Woudt for sending along the SAAO
time-series data, and to Joe Patterson and Rob Robinson for 
useful discussions.  An anonymous referee caught several 
small errors and offered helpful suggestions.  This publication 
makes use of data products from the Two Micron All Sky Survey, 
which is a joint project of the University of
Massachusetts and the Infrared Processing and Analysis Center/California
Institute of Technology, funded by the National Aeronautics and Space
Administration and the National Science Foundation.

\clearpage

\clearpage

\begin{deluxetable}{lrcc}
\tablewidth{0pt}
\tablecolumns{4}
\tablecaption{Journal of Observations}
\tablehead{
\colhead{Date} &
\colhead{$N$} &
\colhead{HA (start)}  &
\colhead{HA (end)} \\
\colhead{[UT]}  &
 &
\colhead{[hh:mm]} &
\colhead{[hh:mm]} \\
}
\startdata
\sidehead{Spectroscopy (2.4 m)}
2003 Jan 31 &  1 & $ -0:13$ & $ -0:13$ \\ 
2004 Jan 18 &  4 & $ +0:17$ & $ +1:30$ \\ 
2004 Jan 19 & 11 & $ -3:52$ & $ +3:06$ \\ 
2004 Jan 20 &  1 & $ +2:38$ & $ +2:38$ \\ 
2004 Mar 8 &  1 & $ -4:07$ & $ -4:07$ \\ 
2004 Mar 9 & 12 & $ -3:59$ & $ +1:32$ \\ 
2004 Nov 18 &  3 & $ -1:22$ & $ -0:58$ \\ 
\sidehead{Filter photometry (2.4m)}
2004 Feb 29 & 12 & $ -1:07$ & $ -0:41$ \\ 
\sidehead{$I$ time-series photometry (2.4m)}
2004 Feb 29 &  5 & $ -0:37$ & $ +4:00$ \\ 
2004 Mar 1 & 23 & $ -3:15$ & $ +4:15$ \\ 
\sidehead{$BVI$ time-series photometry (1.3m)}
2004 May 12 & 42 & $ +0:50$ & $ +4:10$ \\ 
2004 May 16 & 38 & $ +1:16$ & $ +4:13$ \\ 
2004 May 20 & 39 & $ +1:15$ & $ +4:17$ \\ 
\enddata
\end{deluxetable}

\clearpage

\begin{deluxetable}{llrrrr}
\tablewidth{0pt}
\tablecolumns{6}
\tablecaption{Filter Photometry}
\tablehead{
\colhead{$\alpha$\tablenotemark{a}} &
\colhead{$\delta$\tablenotemark{a}} &
\colhead{$U-B$} &
\colhead{$B-V$} &
\colhead{$V$} &
\colhead{$V-I$} \\
}
\startdata
\cutinhead{Field stars}
$10:23:39.24$ &$+00:40:49.4$ & $ -0.32$  & $  0.46$  & $ 19.20$  & $  0.63$  \\ 
$10:23:40.40$ &$+00:39:29.2$ &  \nodata & $  0.98$  & $ 18.59$  & $  1.14$  \\ 
$10:23:43.31$ &$+00:38:19.5$ & $  0.12$  & $  0.69$  & $ 14.86$  & $  0.80$  \\ 
$10:23:47.03$ &$+00:39:08.7$ &  \nodata & $  1.20$  & $ 17.69$  & $  1.39$  \\ 
$10:23:50.53$ &$+00:38:16.6$ &  \nodata & $  1.54$  & $ 19.20$  & $  2.43$  \\ 
$10:23:50.69$ &$+00:37:38.7$ & $ -0.03$  & $  0.58$  & $ 17.40$  & $  0.68$  \\ 
$10:23:50.90$ &$+00:37:16.7$ &  \nodata & $  1.17$  & $ 19.06$  & $  1.41$  \\ 
$10:23:55.67$ &$+00:39:27.3$ &  \nodata & $  1.27$  & $ 18.76$  & $  1.46$  \\ 
$10:23:56.13$ &$+00:36:52.3$ & $  0.37$  & $  0.86$  & $ 16.56$  & $  1.02$  \\ 
\cutinhead{Variable star}
$10:23:47.70$ &$+00:38:41.2$ & $  0.31$  & $  0.69$  & $ 17.46:$  & $  0.78$ \\ 
\enddata
\tablenotetext{a}{Coordinates referred to the ICRS, and are from a fit to 11 
USNO A2.0 stars, with a scatter of 0.4 arcsec.  Units are hours,
minutes, and seconds for the right ascension and degrees, minutes, and 
seconds for the declination.}
\end{deluxetable}

\clearpage

\begin{deluxetable}{lrclrc}
\tablewidth{0pt}
\tabletypesize{\small}
\tablecolumns{6}
\tablecaption{Radial Velocities}
\tablehead{
\colhead{Time\tablenotemark{a}} &
\colhead{$v_{\rm abs}$} &
\colhead{$\sigma$} & 
\colhead{Time\tablenotemark{a}} &
\colhead{$v_{\rm abs}$} &
\colhead{$\sigma$} \\ 
\colhead{} &
\colhead{(km s$^{-1}$)} &
\colhead{(km s$^{-1}$)} &
\colhead{} &
\colhead{(km s$^{-1}$)} &
\colhead{(km s$^{-1}$)} \\
}
\startdata
52670.8778  & $  211$ & $  16$ & 53024.0480  & $  245$ & $  10$ \\
53022.9338  & $ -111$ & $  12$ & 53025.0261  & $  173$ & $  14$ \\
53022.9722  & $ -274$ & $  12$ & 53072.6158  & $  173$ & $  29$ \\
53022.9784  & $ -275$ & $  13$ & 53073.6184  & $  127$ & $  12$ \\
53022.9845  & $ -234$ & $  10$ & 53073.6294  & $   18$ & $  12$ \\
53023.7589  & $ -266$ & $  12$ & 53073.6404  & $  -77$ & $  12$ \\
53023.7678  & $ -284$ & $  10$ & 53073.6514  & $ -165$ & $  11$ \\
53023.8092  & $   -5$ & $  10$ & 53073.6644  & $ -230$ & $  12$ \\
53023.8522  & $  252$ & $   9$ & 53073.6754  & $ -265$ & $  10$ \\
53023.8944  & $  144$ & $  10$ & 53073.7901  & $  282$ & $  13$ \\
53023.9037  & $   71$ & $  11$ & 53073.8012  & $  232$ & $  17$ \\
53023.9351  & $ -160$ & $   9$ & 53073.8122  & $  131$ & $  16$ \\
53023.9648  & $ -257$ & $  10$ & 53073.8232  & $   63$ & $  18$ \\
53023.9769  & $ -230$ & $   9$ & 53073.8370  & $  -43$ & $  15$ \\
53024.0391  & $  209$ & $   9$ & 53073.8478  & $ -115$ & $  15$ \\
\enddata
\tablenotetext{a}{Heliocentric Julian date of 
mid-exposure, minus 2 400 000.}
\tablecomments{Absorption radial velocities.  
The uncertainties are from the 
cross-correlation routine.}
\end{deluxetable}


\begin{deluxetable}{llrrcc}
\tablecolumns{6}
\tabletypesize{\small}
\tablewidth{0pt}
\tablecaption{Fit to the Radial Velocities}
\tablehead{
\colhead{$T_0$\tablenotemark{a}} & 
\colhead{$P$} &
\colhead{$K$} & 
\colhead{$\gamma$} & 
\colhead{$N$} &
\colhead{$\sigma$\tablenotemark{b}}  \\ 
\colhead{} &
\colhead{(d)} & 
\colhead{(km s$^{-1}$)} &
\colhead{(km s$^{-1}$)} & 
\colhead{} &
\colhead{(km s$^{-1}$)} \\
}
\startdata
53025.0008(5) & 0.198094(2) &  268(4) & $ 1(3)$ & 30 &  14 \\ 
\enddata
\tablecomments{Parameters of least-squares sinusoid fits to the radial
velocities, of the form $v(t) = \gamma + K \sin(2 \pi(t - T_0)/P$.}
\tablenotetext{a}{Heliocentric Julian Date minus 2452000.}
\tablenotetext{b}{Root-mean-square residual of the fit.}
\end{deluxetable}

\clearpage

\begin{deluxetable}{cccccccc}
\tablecolumns{8}
\tabletypesize{\scriptsize}
\tablewidth{0pt}
\tablecaption{Selected Light Curve Fit Parameters}
\tablehead{
\colhead{$M_1$} & 
\colhead{$M_2$} &
\colhead{$i$} & 
\colhead{$K_2$} & 
\colhead{$K_2$(back)} &
\colhead{$L_1$} &  
\colhead{$T_{\rm eff (0)}$} &  
\colhead{$d$} \\ 
\colhead{(M$_{\odot}$)} &
\colhead{(M$_{\odot}$)} &
\colhead{(deg)} & 
\colhead{(km s$^{-1}$)} &
\colhead{(km s$^{-1}$)} & 
\colhead{(L$_{\odot}$)} &
\colhead{(kelvin)} & 
\colhead{(pc)} \\
}
\startdata
\cutinhead{$x = 0.5, \alpha = 0.08$}
2.00 & 0.20 & 34.0 & 241 & 260 & 2.63 & 5648 & 1281 \\
2.00 & 0.50 & 37.3 & 240 & 263 & 2.48 & 5679 & 1726 \\
2.00 & 0.70 & 39.1 & 238 & 264 & 2.45 & 5692 & 1934 \\
2.00 & 0.90 & 42.5 & 243 & 273 & 2.37 & 5707 & 2108 \\
2.00 & 1.50 & 48.6 & 238 & 272 & 2.34 & 5728 & 2532 \\
2.80 & 0.10 & 31.0 & 259 & 274 & 3.65 & 5609 & 1035 \\
2.80 & 0.30 & 34.1 & 270 & 291 & 3.29 & 5650 & 1465 \\
2.80 & 0.50 & 35.9 & 270 & 294 & 3.17 & 5668 & 1727 \\
2.80 & 0.70 & 37.4 & 269 & 295 & 3.10 & 5680 & 1930 \\
2.80 & 0.90 & 38.5 & 266 & 295 & 3.08 & 5687 & 2101 \\
2.80 & 1.50 & 44.6 & 272 & 309 & 2.93 & 5715 & 2507 \\
\cutinhead{$x = 0.0, \alpha = 0.08$}
2.00 & 0.20 & 34.2 & 243 & 261 & 2.98 & 5604 & 1268 \\
2.00 & 0.50 & 36.6 & 236 & 259 & 2.84 & 5638 & 1712 \\
2.00 & 0.70 & 37.8 & 231 & 256 & 2.82 & 5651 & 1921 \\
2.00 & 1.50 & 48.0 & 236 & 270 & 2.57 & 5708 & 2524 \\
2.80 & 0.10 & 32.8 & 273 & 289 & 3.92 & 5577 & 1022 \\
2.80 & 0.30 & 34.5 & 272 & 294 & 3.70 & 5607 & 1450 \\
2.80 & 0.50 & 35.7 & 269 & 292 & 3.60 & 5626 & 1712 \\
2.80 & 0.70 & 36.7 & 265 & 290 & 3.56 & 5638 & 1915 \\
2.80 & 0.90 & 37.7 & 262 & 289 & 3.52 & 5649 & 2087 \\
2.80 & 1.50 & 45.8 & 277 & 315 & 3.17 & 5698 & 2493 \\
\cutinhead{$x = 0.5, \alpha = 0.00$}
1.00 & 0.70 & 61.1 & 224 & 259 & 1.19 & 5733 & 1929 \\
1.40 & 0.20 & 39.5 & 238 & 259 & 1.72 & 5684 & 1267 \\
1.40 & 0.30 & 43.6 & 248 & 271 & 1.61 & 5702 & 1442 \\
1.40 & 0.50 & 48.3 & 249 & 278 & 1.56 & 5716 & 1709 \\
2.00 & 0.10 & 36.0 & 262 & 279 & 2.38 & 5655 & 1023 \\
2.00 & 0.20 & 38.1 & 267 & 288 & 2.24 & 5674 & 1273 \\
2.00 & 0.30 & 39.7 & 268 & 291 & 2.17 & 5685 & 1449 \\
2.00 & 0.50 & 44.7 & 279 & 307 & 2.03 & 5706 & 1709 \\
2.00 & 0.70 & 48.0 & 280 & 312 & 1.98 & 5715 & 1912 \\
2.00 & 0.90 & 50.9 & 279 & 315 & 1.96 & 5722 & 2084 \\
2.00 & 1.20 & 55.9 & 279 & 319 & 1.94 & 5728 & 2305 \\
\cutinhead{$x = 0.75, \alpha = 0.08$}
2.00 & 0.30 & 34.7 & 239 & 258 & 2.38 & 5681 & 1467 \\
2.00 & 0.70 & 39.1 & 238 & 264 & 2.27 & 5709 & 1942 \\
2.00 & 1.20 & 44.6 & 236 & 269 & 2.24 & 5728 & 2344 \\
2.00 & 1.50 & 47.2 & 233 & 266 & 2.25 & 5736 & 2542 \\
2.80 & 0.10 & 30.4 & 255 & 270 & 3.40 & 5634 & 1043 \\
2.80 & 0.30 & 33.5 & 266 & 286 & 3.06 & 5669 & 1474 \\
2.80 & 0.50 & 35.4 & 268 & 290 & 2.94 & 5686 & 1737 \\
2.80 & 0.70 & 37.1 & 268 & 293 & 2.88 & 5698 & 1940 \\
2.80 & 0.90 & 38.5 & 266 & 294 & 2.85 & 5706 & 2111 \\
2.80 & 1.50 & 43.4 & 266 & 302 & 2.80 & 5725 & 2518 \\
\cutinhead{$x = 0.50, \alpha = 0.16$}
2.80 & 0.10 & 29.0 & 244 & 257 & 3.93 & 5573 & 1036 \\
2.80 & 0.30 & 31.0 & 248 & 266 & 3.66 & 5608 & 1468 \\
2.80 & 0.50 & 32.8 & 250 & 270 & 3.50 & 5633 & 1732 \\
2.80 & 0.70 & 34.3 & 250 & 272 & 3.42 & 5649 & 1936 \\
2.80 & 0.90 & 35.3 & 247 & 272 & 3.39 & 5659 & 2108 \\
2.80 & 1.50 & 38.1 & 239 & 270 & 3.37 & 5682 & 2524 \\
\enddata
\end{deluxetable}

\clearpage

\begin{figure}
\epsscale{0.88}
\caption{Finding chart for FIRST 1023+00 from the median of three
$V$-band images taken 2004 March.  Stars are labeled with their
$V$ magnitudes, and FIRST 1023+00 is indicated by the two
short lines.  Scale and orientation are as indicated.
[This large figure has been converted to jpg format for the 
astro-ph submission.  It should be available as a separate file.]
}
\end{figure}

\begin{figure}
\epsscale{0.88}
\plotone{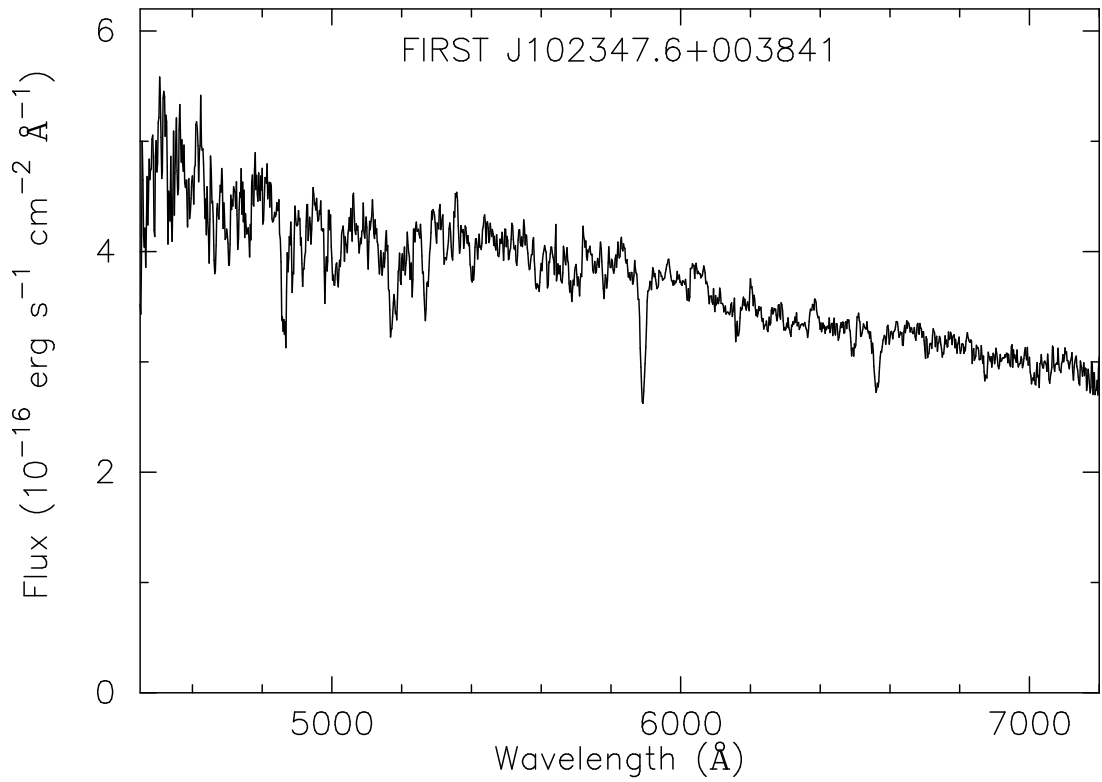}
\caption{Mean spectrum of FIRST 1023+00.  
}
\end{figure}

\begin{figure}
\epsscale{1.0}
\plotone{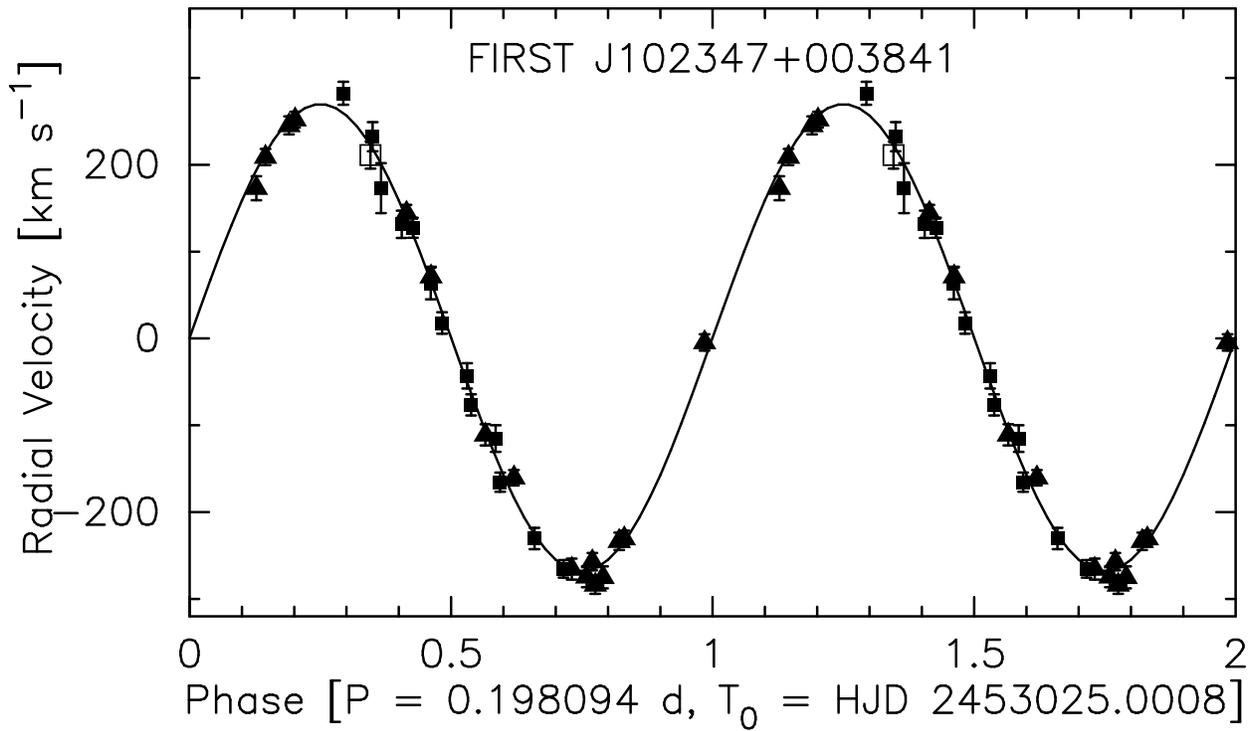}
\caption{Absorption
Radial velocities folded on the best-fit period, with the 
best-fit sinusoid superposed.  All data are plotted twice 
for continuity.  Solid triangles are from 
2004 January, solid squares from 2004 March, and the open square 
is the single
point from 2003 January.  
}
\end{figure}

\begin{figure}
\epsscale{0.9}
\plotone{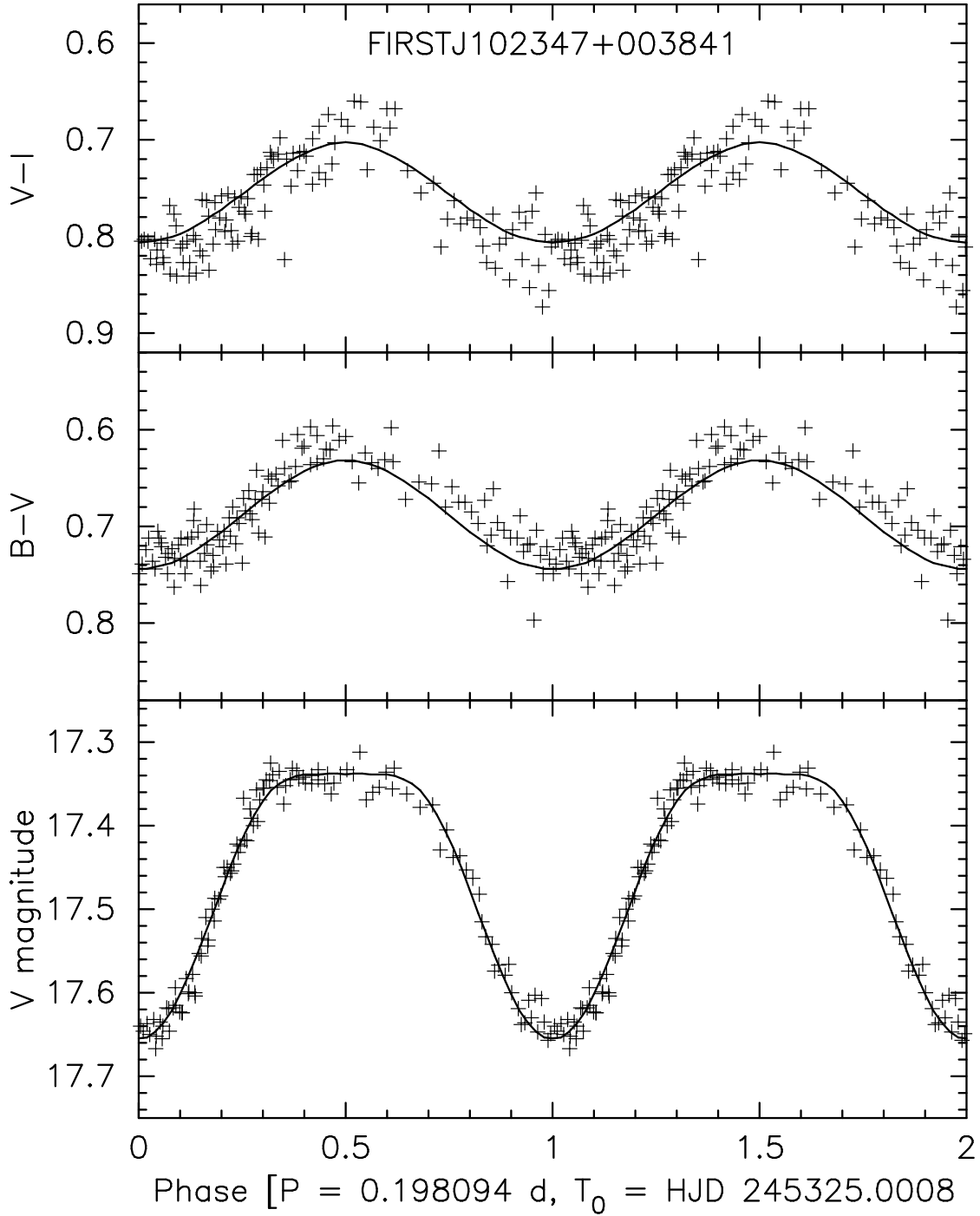}
\caption{Time series photometry from 2004 May.  
The vertical scales are 
adjusted to match the filter photometry of the field.
The horizontal axis gives phase in the {\it spectroscopic}
ephemeris, so zero phase corresponds to inferior conjunction
of the absorption-line source.  All points are shown
twice for continuity.  The solid curve is calculated from the
heating-effect model discussed in the text; this fit is
for $M_1 = 2$ M$_{\odot}$, $M_2 = 0.7$ M$_{\odot}$,
$i = 39$ degrees,  
$T_{\rm eff,0} = 5691$ kelvin, and $L_1 = 2.45$ L$_{\odot}$.
}
\end{figure}

\clearpage

\begin{figure}
\epsscale{1.0}
\caption{
Greyscale representation of the spectra from 2004 March,
arranged according to orbital phase.  The data are repeated
to maintain phase continuity.  The greyscale is positive (dark = 
absorption), and no emission lines are visible.  The phase
coverage is non-uniform because of weather.
[This large figure has been converted to jpg format for the 
astro-ph submission.  It should be available as a separate file.]
}
\end{figure}

\clearpage

\begin{figure}
\epsscale{0.7}
\plotone{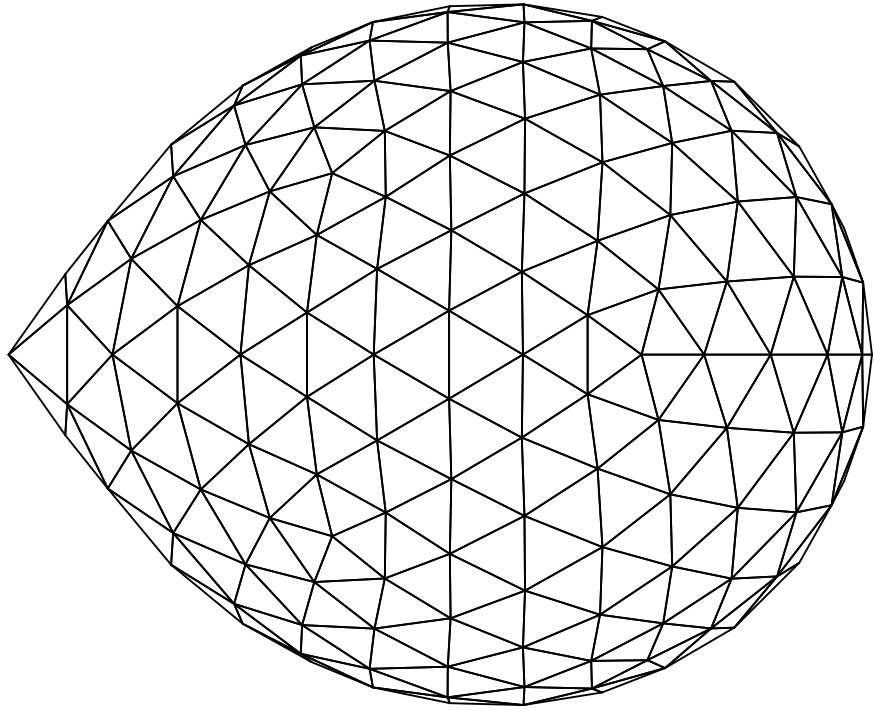}
\caption{
The icosahedral tessellation of the secondary used in the light-curve
modeling.  This view is perpendicular to the orbit plane, and corresponds
to a mass ratio $q = M_2 / M_{\rm wd} = 2/3$.
}
\end{figure}

\begin{figure}
\epsscale{0.9}
\plotone{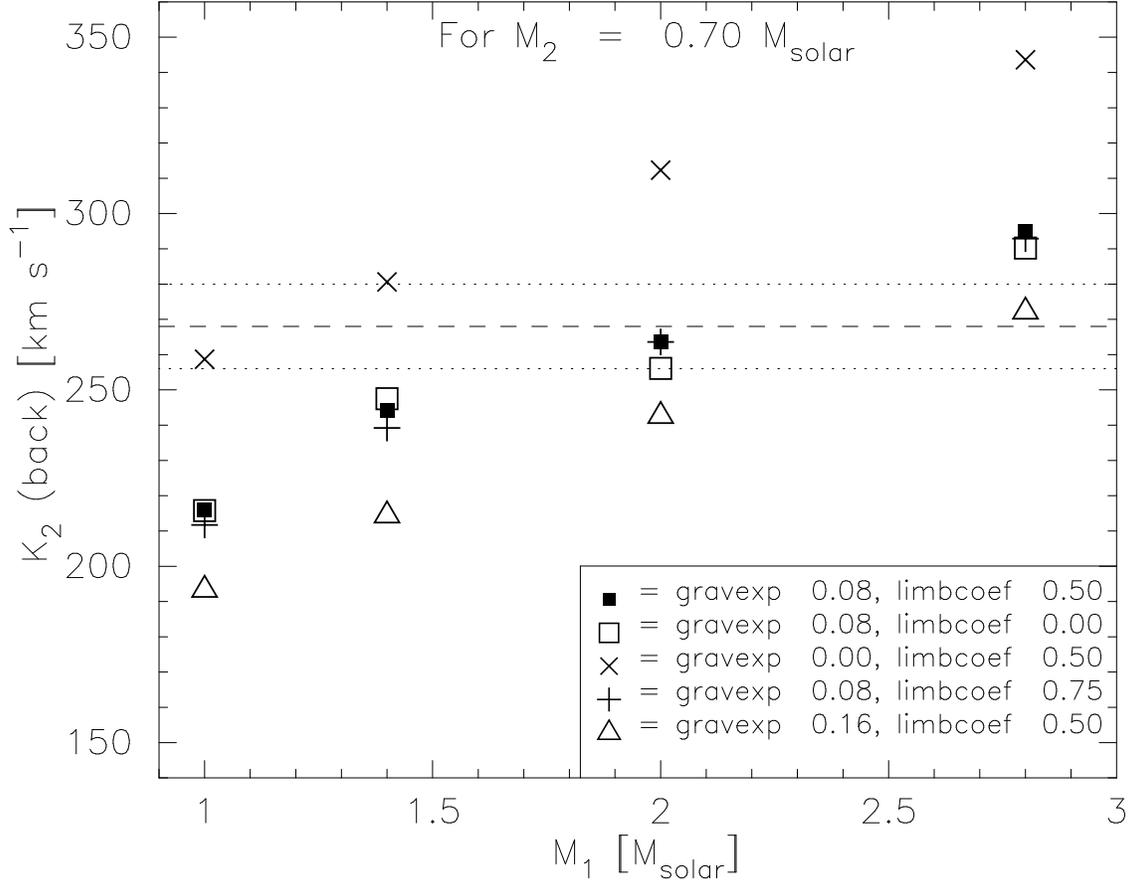}
\caption{
Some results from fitting models to the light curves.  
The velocity amplitude $K_2$(back), computed for the unilluminated portion of the
secondary star, is plotted against the primary mass $M_1$ in solar masses.
Computations for a secondary mass $M_2 = 0.7 {\rm M}_{\odot}$ are shown, but 
this choice makes little difference to the outcome (see text).
The symbols represent choices for limb- and gravity-darkening; the horizontal
dashed line is the observed $K_2$, and the dotted lines are drawn at 
$\pm 3$ standard deviations.  Note that primary star masses 
below the Chandrasekhar limit are possible in the context of the models, but 
require low gravity darkening and the suppression of the radial-velocity signal 
from illuminated portions of the secondary.
}
\end{figure}

\end{document}